\documentstyle[preprint,eqsecnum,aps]{revtex}
 \tightenlines

\newtheorem{theorem}{Theorem}

\newtheorem{corollary}[theorem]{Corollary}

\newtheorem{lemma}[theorem]{Lemma}

\newcommand{\be}{\begin{equation}}
\newcommand{\ee}{\end{equation}}
\newcommand{\bea}{\begin{eqnarray}}
\newcommand{\eea}{\end{eqnarray}}
\newcommand{\ba}{\begin{array}}
\newcommand{\ea}{\end{array}}

\newcommand{\Th}{\Theta}

\newcommand{\la}{\lambda}

\newcommand{\de}{\delta}

\newcommand{\pa}{\partial}

\newcommand{\no}{\nonumber}

\newcommand{\res}{\mbox{res}}

\begin{document}
 \draft

\title
{\sc On the  Miura map between the dispersionless KP and
dispersionless modified KP hierarchies\/}
\author{Jen-Hsu Chang$^1$
 and Ming-Hsien Tu$^2$\\ $^1$Institute of Mathematics, Academia Sinica,\\
Nankang, Taipei, Taiwan\\ E-mail: changjen@math.sinica.edu.tw\\
$^2$Department of Physics, National Chung Cheng University,
\\ Minghsiung, Chiayi, Taiwan\\
E-mail: phymhtu@ccunix.ccu.edu.tw}
\date{\today}
\maketitle
\begin{abstract}
We investigate the Miura map between the dispersionless KP and
dispersionless modified KP hierarchies. We show that the Miura map
is canonical with respect to their bi-Hamiltonian structures.
Moreover, inspired by the works of Takasaki and Takebe, the
twistor construction  of solution structure for the dispersionless
modified KP hierarchy is given.
\end{abstract}

\newpage

 \section{Introduction}

The dispersionless KP hierarchy(dKP)
\cite{Car1,Car2,KG,LM,Ta2,Ta1} can be thought as the
semi-classical limit of the KP hierarchy \cite{Di}. There are many
mathematical and physical problems associated with the dKP
hierarchy and its various reductions, such as Whitham hierarchy,
topological field theory and its connections to string theory and
2D gravity \cite{Car1,Ak,Du4,du1,kr1,kr2}. Similarly, the
dispersionless modified KP(dmKP) hierarchy \cite{Li} can be
regarded as the semi-classical limit of the modified KP (mKP)
hierarchy \cite{jm,ku1} . However, in contrast to dKP, the
integrable structures of dmKP are less investigated. This motives
us to study the relationships between dKP and dmKP and to gain an
insight of dmKP from dKP.

The Miura map \cite{Mi} has been playing an important role in the
development of soliton theory. It's a transformation between two
nonlinear equations, which in general cannot be solved easily.
However, knowing the solutions of one of the non-linear systems,
one may obtain the solutions of the other one via an appropriate
Miura map. A typical example is the Miura map between the KP
equation and the mKP equation \cite{ki,KO,ku2,OR,St1,ST}.
Motivated by the Miura map between the KP equation and the mKP
equation, we will construct the  Miura map between dKP and dmKP.
(In \cite{Ku3}, this Miura map is constructed in different way.)
Moreover, since almost all the known integrable systems are
Hamiltonian, exploring the Hamiltonian nature of these Miura maps
will deepen our understanding of these  relations between these
integrable systems.

 Recently, the canonical property of the
Miura map between the mKP and the KP hierarchy has been
investigated \cite{ST}. It turns out that the Miura map is a
canonical map in the sense that the first and second Hamiltonian
structures of the mKP hierarchy  \cite{O,OS} are mapped to the
first and second Hamiltonian structures of the KP hierarchy. Since
the bi-Hamiltonian structures of mKP and KP have their own
correspondences in dmKP and dKP, thus we expect that the
bi-Hamiltonian structures of dKP and dmKP are still preserved
under the Miura map. We will show, in section 4, that it is indeed
the case.

On the other hand, the solution structure of dKP is also an
interesting subject. To extend the tau-function theory in KP
theory to the semi-classical one in dKP hierarchy and
dispersionless analogue of Virasoro constraints \cite{kr1},
Takasaki and Takebe \cite{Ta2} proposed twistor construction of
the dKP hierarchy using the Orlov function, which can be regarded
as the semi-classical limit of the Orlov operator in KP theory
\cite{Or1,VM}. Using the Miura map between dKP and dmKP, we can
construct the Orlov function of the dmKP hierarchy and hence
establish the twistor construction for dmKP. \\ \indent Our paper
is organized as follows: Section II is background materials for
dKP and dmKP; Section III is the Miura map between dKP and dmKP;
Section IV proves the canonical property of the Miura map; Section
V shows the twistor construction of the dmKP hierarchy; Section VI
lists some unsolved problems.

\section{background materials}

\subsection{dKP hierarchy}

Let's start with the KP hierarchy. The Lax operator of the KP
hierarchy is ($\partial=\partial_x$)
 \begin{eqnarray}
 L=\partial+ \sum_{n=1}^{\infty}u_{n+1}\partial^{-n} \no
\end{eqnarray}
and the KP hierarchy is determined by the Lax equations
($\partial_n=\frac{\partial}{\partial t_n},t_1=x$)
\begin{eqnarray}
\partial_n L=[B_n, L], \label{lax}
\end{eqnarray}
where $B_n=(L^n)_+$ is the differential part of $L^n$. The Lax
equation (\ref{lax}) is equivalent to the existence of the wave
function $\Psi_{KP}$ such that
\begin{eqnarray}
L \Psi_{KP} &=&\lambda \Psi_{KP}, \no \\
\partial_n \Psi_{KP} &=& B_n \Psi_{KP}.\no \label{line}
\end{eqnarray}
Now for the dKP hierarchy, one can think of fast and slow
variables or averaging procedures, by simply taking $t_n \to
\epsilon t_n=T_n$($t_1=x, \epsilon x=X$) in the KP equation
\be
u_t=\frac{1}{4}u_{xxx}+3uu_x+\frac{3}{4} \partial_x^{-1}u_{yy},
\qquad (y=t_2,t=t_3)
 \label{kp}
 \ee
  with $\partial_n \to \epsilon
\frac{\partial}{\partial T_n}$ and $u(t_n) \to U(T_n)$ to obtain
\begin{eqnarray}
\partial_T U=3UU_X+\frac{3}{4}\partial_X^{-1}U_{YY} \label{dkp}
\end{eqnarray}
when  $\epsilon \to 0$ and thus the dispersionless term $u_{xxx}$
is removed. In terms of hierarchies we write
\begin{eqnarray}
L_{\epsilon}=\epsilon \partial +
\sum_{n=1}^{\infty}u_{n+1}(T/\epsilon) (\epsilon \partial)^{-n}
\no
\end{eqnarray}
and think of $u_n(T/\epsilon)=U_n(T)+O(\epsilon)$, etc. One then
takes a WKB form for the wave function $\Psi_{KP}$ with the action
$S_{KP}$
\begin{eqnarray}
\Psi_{KP}=\exp[\frac{1}{\epsilon}S_{KP}(T,\lambda)] \no.
\end{eqnarray}
Now, we replace $\partial_n $ by $\epsilon
\frac{\partial}{\partial T_n}$ and define $P=\partial_X S_{KP}$.
Then $\epsilon^i \partial^i \Psi_{KP} \to P^i \Psi_{KP}$ as
$\epsilon \to 0$ and  the equation $L \Psi_{KP}= \lambda
\Psi_{KP}$ implies
\[
\lambda=P+\sum_{n=1}^{\infty}U_{n+1}(T)P^{-n}.
\]
We also note from $\partial_n \Psi_{KP}=B_n \Psi_{KP}$ that one
obtains $\frac{\partial S_{KP}}{\partial T_n}={{\mathcal B}}_n
(P)=(\lambda^n)_+$, where the subscript $(+)$ now refers to powers
of $P$. The KP hierarchy goes to
\begin{equation}
\frac{\partial P}{\partial T_n}=\frac{ \partial {\mathcal
B}_n(P)}{\partial X}. \label{kph}
\end{equation}

Also, the Lax equation (\ref{lax}) goes to
\begin{equation}
\partial_n \lambda =\{{\mathcal B}_n (P), \lambda \}, \label{zero}
\end{equation}
where the Poisson bracket $\{,\}$ is defined by

\begin{equation}
\{f(X,P),g(X,P)\}=\frac{\partial f}{\partial P} \frac{\partial
g}{\partial X}- \frac {\partial f}{\partial X}\frac{\partial
g}{\partial P}. \label{poss}
\end{equation}
Notice that both the equations (\ref{kph}) and (\ref{zero}) are
compatible respectively, i.e, $\partial^2 \lambda / \partial T_n
\partial T_m= \partial^2 \lambda / \partial T_m \partial T_n$ ,
$\partial^2 P / \partial T_n \partial T_m= \partial^2 P / \partial
T_m \partial T_n$, and they both  imply the dKP hierarchy
\begin{equation}
\frac{\partial {\mathcal B}_n(P)}{\partial T_m}-\frac {\partial
{\mathcal B}_m(P)}{\partial T_n} + \{{\mathcal B}_n(P), {\mathcal
B}_m(P) \}=0.
\end{equation}

In particular,
 \begin{eqnarray}
{\mathcal B}_2(P) &= &P^2+2U_2, \no \\ {\mathcal B}_3 (P)  &=&
P^3+ 3U_2P +3U_3. \no
 \end{eqnarray}
 Then ($T_2=Y, T_3=T$)
\[
\frac{\partial {\mathcal B}_2(P)}{\partial T}-\frac {\partial
{\mathcal B}_3(P)}{\partial Y} + \{{\mathcal B}_2(P), {\mathcal
B}_3(P) \}=0
\]
becomes
\begin{eqnarray*}
U_{3X}&=& \frac{1}{2}U_{2Y}, \\ U_{3Y} &= & \frac{2}{3}
U_{2T}-2U_2U_{2X}
\end{eqnarray*}
and thus
\[
\frac{1}{2}U_{2YY}=\frac{2}{3}(U_{2T}-3U_2U_{2X})_X.
\]
This is the dKP equation (\ref{dkp}) ($U_2=U$).

In summary, we define the dKP hierarchy by
\begin{eqnarray}
\lambda &=& P+\frac{U_2}{P}+\frac{U_3}{P^2}+\cdots, \label{expr}
\\
\partial_n \lambda &=&\{{\mathcal B}_n (P), \lambda \}. \label{zero1}
\end{eqnarray}
 Let us define the Hamiltonians $H_k= 1/k  \int  \res
(\lambda^k),$ where $\res$ means the coefficient of $P^{-1}$, then
the bi-Hamiltonian structure of dKP (\ref{zero1}) is given by
 ~\cite{FR,Li}
\[
\frac{\partial \lambda}{\partial T_k} = \{H_k, \lambda
\}=\Theta^{(2)}(dH_k)=\Theta^{(1)} (dH_{k+1}), \qquad k=1,2,
\cdots
\]
where the Hamiltonian one-form $dH_k$ and the Hamiltonian maps
$\Th^{(i)}$ are defined by
\begin{eqnarray}
 dH_k &=&  \frac{\delta H_k}{\delta U_2}+ \frac{\delta
H_k}{\delta U_3}P+ \frac{\delta H_k}{\delta U_4}P^2 + \frac{\delta
H_k}{\delta U_5}P^3+ \cdots, \no\\
 \Theta^{(2)}(dH_k) &=& \lambda \{\lambda, dH_k \}_+-\{\lambda
,(\lambda dH_k)_+ \} \label{ham1}
 \\ &&+ \{\lambda, \int^X  \res
\{\lambda, dH_k \} \}, \no  \\ \Theta^{(1)}(dH_{k+1}) &=&
\{\lambda, dH_{k+1} \}_+ -\{\lambda ,(dH_{k+1})_+ \} ,\no
\label{ham2}
\end{eqnarray}
 the third term of (\ref{ham1}) being Dirac reduction for
$U_1=0$. \\

\subsection{dmKP}

The Lax operator of the mKP hierarchy is defined \cite{KO} by

 \[
K=\partial +v_0 + v_1 \partial^{-1} +v_2 \partial^{-2} + \cdots.
 \]
which satisfies the Lax equations
\begin{equation}
\partial_n K =[Q_n, K] \label{mkp},
\end{equation}
where $Q_n=(K^n)_{\geq 1}$ means the part of order $\geq 1$ of
$K^n$. Also, the Lax equation (\ref{mkp}) is equivalent to the
existence of wave function $\Psi_{mKP}$ such that
\begin{eqnarray}
K \Psi_{mKP} &=& \mu \Psi_{mKP},  \no \\
\partial_n \Psi_{mKP} &=& Q_n \Psi_{mKP}.\no \label{lmkp}
\end{eqnarray}

To obtain the dmKP hierarchy, similarly, one takes $t_n \to
\epsilon t_n =T_n(t_1=x \to \epsilon t_1 =X) $ in the mKP equation
\be
\label{mkpeq}
v_t=\frac{1}{4}v_{xxx}-\frac{3}{2}v^2v_x+\frac{3}{2}v_x
\partial_x^{-1}v_y
+\frac{3}{4} \partial_x^{-1}v_{yy},
 \ee
 with $\partial_n \to
\epsilon \partial / \partial T_n$ and $v(t_n) \to V(T_n)$ to get
\begin{equation}
V_T=-\frac{3}{2}V^2V_X+\frac{3}{2}V_X \partial_X^{-1}V_Y
+\frac{3}{4} \partial_X^{-1}V_{YY} \label{dmkp},
\end{equation}
when $ \epsilon \to 0$ . Thus, the dispersionless term $v_{xxx}$
is removed, too. In terms of hierarchies, we write
\[
K_{\epsilon}=\epsilon \partial+ v_{1}(T/\epsilon)(\epsilon
\partial)^{-1}
+ v_{2}(T/\epsilon)(\epsilon \partial)^{-2} +\cdots
\]
and think of $v_n( T/ \epsilon)=V_{n}(T)+0(\epsilon)$. One then
takes a WKB form for the wave function $\Psi_{mKP}$ with the
action $S_{mKP}$:
\[
\Psi_{mKP}= \exp (\frac{1}{\epsilon} S_{mKP}( T, \mu)).
\]
Now we replace $\partial_n$ by $ \epsilon \partial / \partial T_n$
and define $P=\partial_X S_{mKP}$. Then $\epsilon^i \partial^i_X
\Psi_{mKP}  \to P^i \Psi_{mKP}$  as $\epsilon \to 0$ and the
equation $K \Psi_{mkP} =\mu \Psi_{mKP}$  yields
\[
\mu =P+\sum_{n=o}^{\infty}V_{n}(T)P^{-n}.
\]

From $\partial_n \Psi_{mKP}= Q_n \Psi_{mKP}$, one obtains
$\partial S_{mKP}/ \partial T_n = {\mathcal Q}_n(P)= (
\mu^n)_{\geq 1}$, where the subscript $\geq 1$ refers to powers  $
\geq 1$ of $P$. The dmKP hierarchy goes to
\[
\frac{\partial P}{\partial T_n}= \frac{\partial {\mathcal
Q}_n(P)}{\partial X} .
\]

It also can be written as the following zero-curvature form

\[
\frac{\partial {\mathcal Q}_n(P)}{\partial T_m}-\frac {\partial
{\mathcal Q}_m(P)}{\partial T_n} + \{{\mathcal Q}_n(P), {\mathcal
Q}_m(P) \}=0,
\]
where the Poisson bracket is defined by (\ref{poss}). In
particular,
\begin{eqnarray*}
{\mathcal Q}_2(P)& =& P^2+2PV_0, \\ {\mathcal Q}_3(P) &=& P^3
+3P^2 V_0 +P(V_1 +3V_0^2).
\end{eqnarray*}

Then the equation($T_2=Y, T_3=T$)
\[
\frac{\partial {\mathcal Q}_2(P)}{\partial T}-\frac {\partial
{\mathcal Q}_3(P)}{\partial Y} + \{{\mathcal Q}_2(P), {\mathcal
Q}_3(P) \}=0,
\]
becomes
\begin{eqnarray}
V_{1X} &=& \frac{3}{2} V_{0Y}-\frac{3}{2}(V_0^2)_X,  \label{mot1}
\\
V_{1Y} &=& 2 V_{0T}-3V_0V_{0Y}-2V_1V_{0X}. \no
\end{eqnarray}
which implies the dmKP (\ref{dmkp}) ($V_0=V$).

In summary, we write  the dmKP equation as
\begin{eqnarray}
\mu &= &P+V_0+\frac{V_1}{P}+ \frac{V_2}{P^2}+\cdots, \no \\
\partial_n \mu &=& \{ {\mathcal Q}_n(P), \mu \}. \label{dmkp1}
\end{eqnarray}
\indent If we define the Hamiltonians as $H_k=\frac{1}{k} \int
\res(\mu^k)$, then the bi-Hamiltonian structure of (\ref{dmkp1})
is described by ~\cite{Li}
\[
\frac{\partial \mu}{\partial T_k}=\{H_k, \mu
\}=J^{(2)}(dH_k)=J^{(1)}(dH_{k+1})
\]
where
\begin{eqnarray}
dH_k &=& \frac{\delta H_k}{\delta V_0} P^{-1} + \frac{\delta
H_k}{\delta V_1} + \frac{\delta H_k}{\delta V_2} P + \frac{\delta
H_k}{\delta V_3} P^2+ \cdots,  \no \\ J^{(2)}(dH_k) &=& \mu \{\mu
, dH_k \}_{\geq -1}-\{\mu, (\mu dH_k)_{\geq 1} \}, \label{poss1}
\\ J^{(1)}(dH_{k+1}) &=&  \{\mu , dH_{k+1} \}_{\geq -1}-\{\mu,
(dH_{k+1})_{\geq 1} \}. \label{possi}
\end{eqnarray}

\section{Dispersionless Miura Map}

It has been shown \cite{ki,KO,ku2,OR,St1,ST} that there exists a
gauge transformation (Miura map) between the Lax operator $L$  of
KP and the Lax operator $K$ of mKP, namely,
\begin{equation}
K=\Phi^{-1}(t)L \Phi(t) \label{gauge},
\end{equation}
where $\Phi(t)$ is an eigenfunction of L, i.e.,
\begin{equation}
\partial_n \Phi=(L^n)_+ \Phi. \label{eigen}
\end{equation}
One generalizes this result to dispersionless limit case.
 \\
\indent Let
\[
{\mathcal L}=P^m+a_{m-1}P^{m-1}+a_{m-2}P^{m-2}+\cdots+a_0+
\frac{a_{-1}}{P} +\frac{a_{-2}}{P^2}+\cdots,
\]
where $a_{m-1}, a_{m-2}, \cdots,a_0 ,a_{-1}, a_{-2}, \cdots$ are
functions of $T=(T_1=X, T_2, T_3, \cdots)$. Also, we suppose
$\phi(T)$(independent  of $P$) is any function of $T$. We define
\begin{eqnarray*}
\tilde {{\mathcal L}} &=& e^{-ad \phi(T)} {\mathcal L},  \\ & =&
{\mathcal L}-\{\phi, {\mathcal L}\}+\frac{1}{2}\{\phi,
\{\phi,{\mathcal L}\}\} -\frac{1}{3!} \{\phi ,\{\phi,
\{\phi,{\mathcal L}\}\}\}+ \cdots
\end{eqnarray*}
where the Poisson bracket is defined by (\ref{poss}). Since $\phi$
is independent of $P$, a simple calculation gets
\begin{equation}
\tilde {{\mathcal L}} =\sum_{n=0}^{\infty} \frac{1}{n!} (\phi_X)^n
\partial_P^n {\mathcal L} \label{simp}.
\end{equation}
\begin{lemma} Let $ \tilde {{\mathcal L}}$ be defined as above. Then
\[
\tilde {{\mathcal L}}_{\geq 1} =  e^{-ad \phi}({\mathcal L}_{\geq
0})-{\mathcal L}_{\geq 0} \vert_{P=\phi_X},
\]
where
\[
 {\mathcal L}_{\geq 0} \vert_{P=\phi_X}= \phi_{X}^m +a_{m-1} \phi_{X}^{m-1}
+\cdots +a_1 \phi_X +a_0.
\]
\end{lemma}
{\it Proof\/}. From (\ref{simp}), one knows that $\tilde {\mathcal
L}_{\geq 0}$ comes from the polynomial part of ${\mathcal L}$.
 Hence
\begin{eqnarray*}
\tilde {{\mathcal L}}_{\geq 1} &=& \tilde {{\mathcal L}}_{\geq
0}-\tilde {{\mathcal L}}_0,  \\ &=& e^{-ad \phi}({\mathcal
L}_{\geq 0})-e^{-ad \phi}({\mathcal L}_{\geq 0}) \vert_{P=0}.
\end{eqnarray*}

Using (\ref{simp}), one knows
\begin{eqnarray*}
 e^{-ad \phi}({\mathcal L}_{\geq 0}) \vert_{P=0} &=& \sum_{n=0}^{\infty}
  \frac{1}{n!} \phi_{X}^n (\partial_P^n {\mathcal L}_{\geq 0} \vert_{P=0}), \\
&=& \sum_{n=0}^{\infty} \frac{1}{n!} \phi_{X}^n ( a_n n!), \\ &=&
\phi_X^m + a_{m-1} \phi_X^{m-1}+a_{m-2} \phi_X^{m-2} +\cdots +a_1
\phi_X +a_0, \\ &=& {\mathcal L}_{\geq 0} \vert_{P=\phi_X}.
\end{eqnarray*}
This completes the lemma.\qquad $\Box$

\begin{lemma}
$e^{-ad \phi} \{f(T,P), g(T,P) \}=\{e^{-ad \phi} f(T,P), e^{-ad
\phi} g(T,P)\}$.
\end{lemma}
{\it Proof\/}.
\begin{eqnarray*}
r.h.s. &=& \sum_{m,n=0}^{\infty} \frac{1}{m!n!} \{\phi_X^n
\partial_P^n f, \phi_X^m \partial_P^m g \}, \\
&=& \sum_{m,n=0}^{\infty} \frac{1}{m!n!} \phi_X^{m+n}
\{\partial_P^n f, \partial_P^m g \}, \\ &=& \sum_{m=0}^{\infty}
\frac{\phi_X^m}{m!} \sum_{n=0}^m \left(
\begin{array}{c}
 m \\ n
\end{array}
\right ) \{\ \partial_P^n f,  \partial_P^{m-n} g \}, \\ &=& e^{-ad
\phi} \{f,g \}=l.h.s.\qquad \Box
\end{eqnarray*}
\begin{theorem} Let $ \tilde {{\mathcal L}}$ be defined as above. Then
\[
\tilde{{\mathcal L}}_{T_q}-\{(\tilde{{\mathcal L}}^q )_{\geq 1} ,
\tilde {{\mathcal L}} \}=e^{-ad \phi}({\mathcal
L}_{T_q}-\{({\mathcal L}^q )_+ , {\mathcal L} \})-
\{\phi_{T_q}-({\mathcal L}^q )_+ \vert_{P=\phi_X}, \tilde{\mathcal
L} \} ,
\]
where the subscript $T_q$ means  $\partial / \partial T_q$.
\end{theorem}
{\it Proof\/}. Using (\ref{simp}), we have
\begin{eqnarray*}
\frac{\partial \tilde{{\mathcal L}}}{\partial T_q} & = & e^{-ad
\phi} \frac{\partial {\mathcal L}}{\partial T_q}+
\sum_{n=0}^{\infty} \frac{1}{n !}[\frac{\partial}{\partial T_q}
(\phi_X)^n]
\partial_P^n {\mathcal L}, \\
 &=& e^{-ad \phi} \frac{\partial {\mathcal L}}{\partial T_q}+
(\frac{\partial^2 \phi}{\partial T_q \partial X})
\sum_{n=0}^{\infty} \frac{1}{n !} \phi_X^{n} \partial_P^{n+1}
{\mathcal L}, \\ &=& e^{-ad \phi} \frac{\partial {\mathcal
L}}{\partial T_q}-\{\phi_{T_q}, e^{-ad \phi} {\mathcal L} \}.
\end{eqnarray*}
Then, by Lemmas 1 and 2, we have
\begin{eqnarray*}
\tilde{{\mathcal L}}_{T_q}-\{(\tilde{{\mathcal L}}^q )_{\geq 1} ,
\tilde {{\mathcal L}} \} &=& e^{-ad \phi}({\mathcal
L}_{T_q}-\{\phi_{T_q}, {{\mathcal L}}\})- \{e^{-ad \phi}({\mathcal
L}^q)_+-({\mathcal L}^q)_+ \vert_{P=\phi_X},
   e^{-ad \phi} {\mathcal L} \}, \\
  &=& e^{-ad \phi}({\mathcal L}_{T_q}-\{({\mathcal L}^q)_+, {\mathcal L}\})
  + \{ ({\mathcal L}^q)_+ \vert_{P=\phi_X}-\phi_{T_q}, \tilde{{\mathcal L}}\}.
\end{eqnarray*}
This completes the theorem. $\Box$

\begin{corollary} Let
\[
{\mathcal L} =P+
\frac{U_2}{P}+\frac{U_3}{P^2}+\frac{U_4}{P^3}+\cdots
\]
and suppose that $U_i(T)$ satisfy the dKP hierarchy (\ref{zero})
($\lambda = {\mathcal L}$) and $\phi (T)$ satisfies the equation
\begin{equation}
\frac{\partial \phi}{\partial T_n}=({\mathcal L}^n)_+
\vert_{P=\phi_X}. \label{deigen}
\end{equation}
Then $\tilde {{\mathcal L}}=e^{-ad \phi} {\mathcal L}$ will
satisfy the dmKP hierarchy (\ref{dmkp1}) ($\mu=\tilde {\mathcal
L}$).
\end{corollary}
{\it Proof\/}. Obvious. $\Box$ \\ From the corollary, one calls
the map
\begin{equation}
{\mathcal L} \to e^{-ad \phi} {\mathcal L} \label{dmp}
\end{equation}
the dispersionless Miura map between dKP and dmKP. It's because
one can think the map (\ref{dmp}) as the dispersionless limit of
equation (\ref{gauge}) and, moreover, the equation (\ref{deigen})
can be regarded as the dispersionless limit of equation
(\ref{eigen}). As in the case of KP and mKP, the dispersionless
Miura map gives rise to a transformation between dKP and dmKP in
terms of "dispersionless" eigenfunction $\phi(T)$. If one assumes
that
\[
\tilde {\mathcal
L}=P+V_0+\frac{V_{1}}{P}+\frac{V_{2}}{P^2}+\frac{V_{3}}{P^3} +
\cdots,
\]
then, after some calculations, one gets
\begin{eqnarray}
V_0 &=& \phi_X, \no \\ V_{1} &=& U_2, \label{mot2} \\ V_{2} &=&
U_3 +\phi_X U_2,  \no \\ V_{3} &=& U_4 +2 \phi_X U_3 + \phi_X^2
U_2, \no \\ V_{4} &=& U_5 +3 \phi_X U_4 + 3\phi_X^2 U_3+\phi_X^3
U_2, \no \\ & \vdots&  \no \\ V_{n} &=& \sum_{i=0}^{n-1} \left(
\begin{array}{c}
 n-1 \\ i
\end{array}
\right) \phi_X^i U_{n+1-i},\,\ n \geq 1.  \no
\end{eqnarray}
\indent Finally, it is well known that Miura-type transformations
between (\ref{kp}) and (\ref{mkpeq}) are
 \bea
 u_1 &=& \frac{3}{2} (-v^2-v_x+\partial_x^{-1}v_y), \no \\
 u_2&=&\frac{3}{2} (-v^2+v_x+\partial_x^{-1}v_y)\no \label{miura}.
 \eea

In the dispersionless limit, the term $v_x$ is removed and we
obtain the only transformation
\begin{equation}
U=\frac{3}{2}(-V^2+\partial_X^{-1}V_Y). \label{dmir}
\end{equation}
Notice that we can also obtain the equation (\ref{dmir}) from
(\ref{mot1}) and (\ref{mot2}). Furthermore, since the term
corresponding to $v_x$ is removed, this would explain why we
cannot find the auto-B\"acklund transformation of the dKP
hierarchy as one did in the ordinary case \cite{St1}.


\section{canonical property of the Miura map}

Having constructed the dispersionless Miura map between the dKP
hierarchy and the dmKP hierarchy in the Lax formulation, which
provides a connection of solutions associated with dKP and dmKP,
we next would like to investigate the canonical property of the
Miura map. As we have seen that both dKP and dmKP hierarchies
equip a compatible bi-Hamiltonian structure, thus it is quite
natural to ask whether their bi-Hamiltonian structures are still
preserved under the Miura map.

To proceed the discussion, it is convenient to rewrite the
dispersionless Miura map as
\begin{equation}\label{gmap}
  G: \mu(T,P)\to \lambda(T,P)=e^{ad \phi(T)}\mu(T,P)
\end{equation}
where $\lambda$ and $\mu$ are Lax operators of the dKP and dmKP
hierarchies respectively and the function $\phi(T)=\int^XV_0$ is
independent of $P$. In the following, the symbols $A$, $B$ and $C$
will stand for arbitrary Laurent series without further mention.

\begin{lemma}  $e^{-ad \phi(T)}e^{ad\phi(T)}A=A.$
 \end{lemma}

{\it Proof\/}.  By definition,
  \bea
 e^{-ad\phi(T)}e^{ad\phi(T)}A
 &=&e^{-ad\phi(T)}\sum_{n=0}^{\infty}\frac{(-1)^n}{n!}\phi_X^n\pa_P^nA,\no\\
&=&\sum_{m=0}^{\infty}\sum_{n=0}^{\infty}\frac{(-1)^n}{m!n!}
\phi_X^{m+n}\pa_P^{m+n}A,\no\\
&=&\sum_{m=0}^{\infty}\frac{1}{m!}\phi_X^m\pa_P^mA
\sum_{n=0}^{m}(-1)^n \left(
\begin{array}{c}
m \\ n
\end{array}
\right ) =A.\no\qquad \Box
  \eea

  \begin{lemma}
$e^{-ad \phi}(AB)=(e^{-ad \phi} A)(e^{-ad \phi} B)$.
\end{lemma}
{\it Proof\/}.
\begin{eqnarray*}
e^{-ad \phi}(AB) &=& \sum_{n=0}^{\infty}
\frac{\phi_X^n}{n!}\partial_P^n(AB),  \\ &=& \sum_{n=0}^{\infty}
\frac{\phi_X^n}{n!} \left [\sum_{m=0}^n \left(
\begin{array}{c}
n \\m
\end{array}
\right) (\partial_P^m A) (\partial_P^{n-m} B) \right ], \\ &=&
\sum_{n=0}^{\infty}  \left [\sum_{m=0}^n  \frac{\phi_X^{n-m}
\phi_X^m}{m!(n-m)!} (\partial_P^m A) (\partial_P^{n-m} B) \right
],\\
 &=& (\sum_{m=0}^{\infty}   \frac{ \phi_X^m}{m!}
\partial_P^m A) (\sum_{n=0}^{\infty}   \frac{ \phi_X^n}{n!}
\partial_P^n B), \\
&=& (e^{-ad \phi} A)(e^{-ad \phi} B).\qquad \Box
\end{eqnarray*}

 \begin{lemma}
 $\int \res(A\{B,C\})=\int \res(\{A,B\}C)$
 \end{lemma}
 {\it Proof\/}.
 \begin{eqnarray*}
 l.h.s.&=&\int \res \left[A\left(\frac{\pa B}{\pa P}\frac{\pa C}{\pa X}-
 \frac{\pa B}{\pa X}\frac{\pa C}{\pa P}\right)\right],\\
 &=&\int \res\left[-\frac{\pa}{\pa X}\left(A\frac{\pa B}{\pa P}\right)C+
 \frac{\pa}{\pa P}\left(A\frac{\pa B}{\pa X}\right)C\right],\\
 &=&r.h.s.\qquad \Box
 \end{eqnarray*}
 To investigate the canonical property of the Miura map (\ref{gmap}) we
 shall first construct the tangential map between the tangent
 spaces (to which $\de \lambda$ and $\de \mu$ belong) of the corresponding
 phase space manifolds.

\begin{theorem} For the Miura map $G$, the linearized map $G'$ and its
transposed map ${G'}^{\dagger}$ are given by
 \begin{eqnarray}
 G'&:& B\to e^{ad\phi(T)}B+\{\int^Xb_0, \la\},
 \label{lmp} \\
 {G'} ^{\dagger}&:& A\to e^{-ad\phi(T)}A+P^{-1}\int^X\res\{A, \la\}
 \label{ltmp}
 \end{eqnarray}
where $b_0\equiv (B)_0$ and $\dagger$ is the transposed operation
defined by $ \int\res( A G'B)=\int \res(({G'}^{\dagger}A) B)$.
\end{theorem}

 {\it Proof\/}.  Let $B=\de\mu$ be an infinitesimal deformation of the Lax operator
  $\mu$, then under the Miura map $G$ we have
  \begin{eqnarray}
  \mu+B &\to& e^{ad(\phi+\int^X b_0)}(\mu+B),\no\\
  &=&e^{ad\phi}\mu+e^{ad\phi}B+\{\int^X b_0, \la\}+O(B^2).\no
  \end{eqnarray}
which implies the linearized map (\ref{lmp}). On the other hand,
using Lemmas 5-7 and the fact $\res(e^{ad\phi}A)=\res(A)$ we have
 \begin{eqnarray}
  \int \res(AG'B)&=&
  \int \res(A(e^{ad\phi}B))+\int \res(A\{\int^X b_0, \la\}),\no\\
  &=&\int \res((e^{-ad\phi}A)B)+\int b_0\int^X \res\{A, \la\},\no\\
  &=&\int \res((e^{-ad\phi}A)B)+\int \res((P^{-1}\int^X \res\{A,
  \la\})B)\no
 \end{eqnarray}
where we have used integration by part and $b_0=\res(BP^{-1})$ to
reach the last line. Comparing the last line with $\int
\res(({G'}^{\dagger}A) B)$ we obtain (\ref{ltmp}).  $\Box$\\
 Now we are in a position to investigate the canonical property of the
Miura map.
\begin{theorem} The Miura map $G$ maps the bi-Hamiltonian structure of the
dmKP hierarchy given by $J^{(1)}$ and $J^{(2)}$ to the
bi-Hamiltonian structure of the dKP hierarchy given by $\Th^{(1)}$
and $\Th^{(2)}$ respectively, i.e., they are related by
 \begin{eqnarray}
  \Th^{(1)}&=&G' J^{(1)}{G'}^{\dagger},
  \label{cano1}\\
  \Th^{(2)}&=&G' J^{(2)}{G'}^{\dagger}
  \label{cano2}
 \end{eqnarray}
where $G'$ and ${G'}^{\dagger}$ are transformations defined in
Theorem 8.
\end{theorem}
 {\it Proof\/}.  To prove the first structure, let us act the right hand side
 of (\ref{cano1}) on an arbitrary Laurent series $A$, then $G'J^{(1)}({G'}^{\dagger}A)=G'B$
  where
   \begin{eqnarray}
 B&\equiv&J^{(1)}({G'}^{\dagger}A),\no\\
 &=&\{\mu, {G'}^{\dagger}A\}_{\ge -1}-\{\mu, ({G'}^{\dagger}A)_{\ge 1}\},\no\\
 &=&e^{-ad\phi}\left(\{\la, A\}_+-\{\la, A_+\}+\{\la,
 (e^{-ad\phi}A)_0\}
 \label{B1}
 \right),
   \end{eqnarray}
   and thus
   \begin{equation}
 \int^X b_0=\int^X (B)_0=(e^{-ad\phi}A)_0.
 \label{b01}
   \end{equation}
   Substituting (\ref{B1}) and (\ref{b01}) into (\ref{lmp}) we have
   \begin{eqnarray}
 G'J^{(1)}({G'}^{\dagger}A)&=&e^{ad\phi}B+\{(e^{-ad\phi}A)_0,
 \la\},\no\\
 &=&\{\la, A\}_+-\{\la, A_+\}=\Th^{(1)}(A).\no
   \end{eqnarray}
   This completes the first part of the proof.
 For the second Hamiltonian structure, using (\ref{poss1}) and (\ref{ltmp})
 we have
   \begin{eqnarray}
 B&\equiv&J^{(2)}({G'}^{\dagger}A),\no\\
 &=&\{\mu, {G'}^{\dagger}A\}_+\mu-\{\mu, (\mu {G'}^{\dagger}A)_+\}+\{\mu,
 (\mu{G'}^{\dagger}A)_0\}+\mu P^{-1}\res\{\mu, {G'}^{\dagger}A\}
 \label{b2}
    \end{eqnarray}
where each term in (\ref{b2}) can be calculated as follows:
 \begin{eqnarray}
 (1)&=& e^{-ad\phi}(\{\la, A\}_+\la),\no\\
 (2)&=& -e^{-ad\phi}\left(\{\la, (A\la)_+\}+\{\la, \int^X\res\{A, \la\}\}\right),\no\\
 (3)&=& e^{-ad\phi}\left(\{\la, (e^{-ad\phi}(A\la))_0\}+
 \{\la, \int^X\res\{A, \la\}\}\right),\no\\
 (4)&=&0.\no
 \end{eqnarray}
Then
 \begin{eqnarray}
 B&=&(1)+(2)+(3)+(4),\no\\
 &=&e^{-ad\phi}\left(\{\la, A\}_+\la-\{\la, (A\la)_+\}+
 \{\la, (e^{-ad\phi}(A\la))_0\} \right),
 \label{B2}
 \end{eqnarray}
and
 \begin{equation}
\int^X b_0=\left(e^{-ad\phi}(\la A) \right)_0+\int^X\res\{A,
\la\}. \label{b02}
 \end{equation}
Substituting (\ref{B2}) and (\ref{b02}) into (\ref{lmp}) we get
 \[
 G' J^{(2)}({G'}^{\dagger}A)=\la\{\la, A\}_+-\{\la, (\la A)_+\}+\{\la, \int^X\res\{\la,
 A\}\}=\Th^{(2)}(A).
 \]
 This completes the theorem.
 $\Box$

\section{solution structure of dispersionless mKP}

In ~\cite{Ta2,Ta1}, it is shown that the twistor construction
exists for the solution structure of dKP hierarchy. Based on the
dispersionless Miura map described in Section III, we can also
find a similar twistor construction for solution structure of
dmKP. This is the purpose of this section.
\\ \indent First of all, let's recall the twistor construction of
dKP in ~\cite{Ta2,Ta1}. Here we change slightly the symbols used
in those papers. Let's consider the dKP (\ref{zero1}). It can be
shown that there exists a Laurent series $\psi(T,P)$ (dressing
function) such that
\[
\lambda=e^{ad \psi}(P),
\]
where $\psi(T,P)$ has the form
\[
\psi(T,P)=\sum_{n=1}^{\infty}\psi_n(T)P^{-n}.
\]
Such Laurent series $\psi(T,P)$ is not unique up to a constant
Laurent series $\sum_{i=0}^{\infty} c_i P^{-i}$. The Orlov
function of dKP is by definition a formal Laurent series
\cite{Ta2,Ta1}
\[
{\mathcal M}=e^{ad \psi} (\sum_{n=1}^{\infty} n T_n P^{n-1}).
\]

It's convenient to expand ${\mathcal M}$ into a Laurent series of
$\lambda$ as
\begin{equation}
{\mathcal M}=\sum_{n=1}^{\infty} n T_n \lambda^{n-1}+
\sum_{i=1}^{\infty}h_i(T) \lambda^{-i}.  \label{orlov}
\end{equation}
It can be also shown that the series ${\mathcal M}$ satisfies the
Lax equation
\begin{equation}
\frac{\partial {\mathcal M}}{\partial T_n}=\{{\mathcal B}_{n},
{\mathcal M} \} \label{orlov1}
\end{equation}
and the canonical Possion relation
\begin{equation}
\{\lambda , {\mathcal M} \}=1. \label{poss2}
\end{equation}

To get the solution structure of dKP hierarchy, let's consider a
pair of two functions $(f(P,X), g(P,X))$ such that they are
arbitrary holomorphic functions defined in a neighborhood of
$P=\infty$ except at $P=\infty$ itself. Then we have the following
fact(twistor construction of dKP hierarchy). \\
 {\bf Fact:(K.Takasaki and T. Takebe, \cite{Ta2})}
  Suppose \\ (i) $\lambda$
and ${\mathcal M}$ has the form (\ref{expr}) and (\ref{orlov}). \\
 (ii) $f(P,X)$ and $g(P,X)$
described as above satisfy the canonical relation
\begin{equation}
 \{f(P,X), g(P,X) \}=1.  \label{one}
\end{equation}
Then the following functional equations(in $P$)
\begin{equation}
f(\lambda, {\mathcal M})_{\leq -1}=0 \hspace{8mm} g(\lambda ,
{\mathcal M})_{\leq -1}=0 \label{twist}
\end{equation}
will imply equations (\ref{zero1}), (\ref{orlov1}) and
(\ref{poss2}), i.e, the pair $(\lambda, {\mathcal M})$ gives a
solution of dKP hierarchy. We call $(f(P,X),g(P,X))$ the twistor
data of this solution. $\parallel$\\
 \indent
 Conversely, each solution of dKP hierarchy possesses a twistor data
corresponding to the solution, i.e, if $(\lambda, {\mathcal M})$
is a solution of (\ref{zero1}), (\ref{orlov1}) and (\ref{poss2}),
then there exists a pair $(f(P,X),g(P,X))$ which satisfies
(\ref{one}) and (\ref{twist}). In fact, if we let $e^{ad
\psi(T,P)}$ be the dressing operator corresponding to $(\lambda,
{\mathcal M})$, then the twistor data $(f,g)$ of this solution
will be
\begin{eqnarray}
f(P,X)& = & e^{-ad \psi_0(X,P)} P, \no \\ g(P,X)& = & e^{-ad
\psi_0(X,P)} X, \label{pair}
\end{eqnarray}
where $\psi_0(X, P)=\psi(T_1=X, T_2=T_3=T_4= \cdots =0, P)$. \\
\indent Next, we consider the dispersionless Miura map (\ref{dmp})
from  dKP to dmKP. Let us define
\begin{eqnarray}
\mu &=& e^{-ad \phi(T)} \lambda, \no  \\ \tilde {{\mathcal M}} &=&
e^{-ad \phi(T)} {{\mathcal M}}.  \label{me}
\end{eqnarray}
Then $\mu$ satisfies dmKP hierarchy (Theorem 3) and from Lemma 2
we have
\begin{equation}
\{\mu, \tilde {{\mathcal M}} \}=1. \label{poss3}
\end{equation}

Morever, a similar argument of Theorem 3  can also show that
\begin{equation}
\frac{\partial \tilde {{\mathcal M}}}{\partial T_n}=\{{\mathcal
Q}_n(P), \tilde {{\mathcal M}} \}. \label{tim}
\end{equation}

Now, we want to construct a pair of twistor data $(\tilde f(P,X),
\tilde g(P,X))$ corresponding to $\mu$ and $\tilde {{\mathcal M}}$
defined in (\ref{me}).
\begin{theorem} Let $(\lambda, {\mathcal
M})$ be  a solution of (\ref{zero1}), (\ref{orlov1}) and
(\ref{poss2}) and $\mu, \tilde {{\mathcal M}} $ is defined by the
Miura map (\ref{me}). If we define
\begin{eqnarray*}
\tilde f (P,X) &=& e^{-ad \psi_0(X,P)} e^{ad \phi_0(X)}P, \\
\tilde g(P,X) &=& e^{-ad \psi_0(X,P)} e^{ad \phi_0(X)}X=g(P,X),
\end{eqnarray*}
where $\psi_0(X,P)$ is defined in (\ref{pair}) and
$\phi_0(x)=\phi(T_1=X, T_2=T_3=\cdots =0)$, (obviously, we have
$\{\tilde f, \tilde g \}=1$.) then
\begin{eqnarray*}
\tilde f (\mu, \tilde {{\mathcal M}})_{\leq 0} &= & 0, \\ \tilde g
(\mu, \tilde {\mathcal M})_{\leq -1}&= & 0.
\end{eqnarray*}
\end{theorem}
{\it Proof\/}. For convenience, we let $T=0$ mean
$T_2=T_3=T_4=\cdots=0$. Since
\begin{eqnarray*}
\lambda (T=0) &= & e^{ad \psi_0} P, \\ {{\mathcal M}} (T=0) &=&
e^{ad \psi_0} X,
\end{eqnarray*}
then we have
\begin{eqnarray*}
\mu(T=0)&=&e^{-ad \phi_0} \lambda (T=0) =e^{-ad \phi_0}e^{ad
\psi_0}P, \\ \tilde {{\mathcal M}}(T=0)&=&e^{-ad \phi_0} {\mathcal
M} (T=0) = e^{-ad \phi_0}e^{ad \psi_0}X .
\end{eqnarray*}

Therefore, by the Lemma 5 and assumptions, we have
\begin{eqnarray}
\tilde f(\mu(T=0), \tilde {{\mathcal M}}(T=0)) &=& e^{-ad
\phi_0}e^{ad \psi_0} \tilde f(P,X), \no \\ &=& e^{-ad \phi_0}e^{ad
\psi_0}(e^{-ad \psi_0}e^{ad \phi_0}P)=P, \label{init1}  \\ \tilde
g(\mu(T=0), \tilde {\mathcal M}(T=0)) &=& e^{-ad \phi_0}e^{ad
\psi_0} \tilde g(P,X), \no \\ &=& e^{-ad \phi_0}e^{ad
\psi_0}(e^{-ad \psi_0}e^{ad \phi_0}X)=X. \label{init2}\no
\end{eqnarray}

Now, we prove that $\tilde f(\mu, \tilde {\mathcal M})_{\leq
0}=0.$ Since $\mu$ and $\tilde {\mathcal M}$ satisfy  equations
(\ref{dmkp1}) and (\ref{tim}) respectively, we have
\[
\frac{\partial \tilde f (\mu, \tilde {\mathcal M})}{\partial
T_n}=\{{\mathcal Q}_n(P), \tilde f (\mu, \tilde {\mathcal M}) \}.
\]
Using (\ref{init1}), we see that $\partial \tilde f(\mu, \tilde
M)/
\partial T_n \vert_{T=0}$ will only contain powers $\geq 1$ of
$P$. In this way, we can prove, by induction, that $(\partial /
\partial T)^{\alpha} \tilde f(\mu, \tilde {\mathcal M})
\vert_{T=0}$, i.e, coefficients of Taylor expansion at $T=0$, will
only contain powers $\geq 1$ of $P$ for any multi-index $\alpha$.
Thus, we have proved that $\tilde f(\mu, \tilde {\mathcal
M})_{\leq 0}=0.$ As for $\tilde g(\mu, \tilde {\mathcal M})_{\leq
-1}=0$, we notice that the powers of $P$ of $\{{\mathcal Q}_n(P),
X \}$ are $\geq 0$. Then it can be proved in the same way. $\Box$

This theorem shows the possibility of twistor construction for the
solution structure of dmKP without using dispersionless Miura map.
Indeed, we have the following main theorem of this section.

\begin{theorem}
Let
\begin{eqnarray*}
\mu &=& P+V_0+\frac{V_1}{P}+\frac{V_2}{P^2}+\cdots, \\ {\mathcal
M}_{dmkp} &=& \sum_{n=1}^{\infty} nT_n
\mu^{n-1}+\sum_{i=1}^{\infty}S_i(T) \mu^{-i}
\end{eqnarray*}
(${\mathcal M}_{dmkp}$ can be defined as the Orlov function of
dmKP). Suppose that
\begin{equation}
\{f(P,X), g(P,X) \}=1. \label{uni}
\end{equation}

Then the functional equations
\begin{eqnarray}
f(\mu, {\mathcal M}_{dmkp})_{\leq 0} &=& 0,  \no \\ g(\mu,
{\mathcal M}_{dmkp})_{\leq -1} &=& 0  \label{fun}
\end{eqnarray}
can get a solution of
\begin{eqnarray*}
\partial_{T_n} \mu &=& \{{\mathcal Q}_{\geq 1}^n(P), \mu \}, \\
\partial_{T_n} {\mathcal M}_{dmkp} &=& \{{\mathcal Q}_{\geq 1}^n(P), {\mathcal M}_{dmkp} \}, \\
\{ \mu, {\mathcal M}_{dmkp} \} &= & 1.
\end{eqnarray*}
\end{theorem}
{\it Proof\/}. For convenience, we let
\begin{eqnarray}
\tilde \mu &=& f(\mu, {\mathcal M}_{dmkp}), \no \\ \tilde
{\mathcal M}_{dmkp} &=& g(\mu, {\mathcal M}_{dmkp}). \label{sub}
\end{eqnarray}

We first derive the canonical Poisson relation. By differentiating
the last equations with respect to $P$ and $X$, we have
 \be
\left(
\begin{array}{cc}
\frac{\partial f(\mu, {\mathcal M}_{dmkp})}{\partial \mu} &
\frac{\partial f(\mu, {\mathcal M}_{dmkp})}{\partial {\mathcal
M}_{dmkp}} \\ \frac{\partial g(\mu, {\mathcal M}_{dmkp})}{\partial
\mu} & \frac{\partial g(\mu, {\mathcal M}_{dmkp})}{\partial
{\mathcal M}_{dmkp}}
\end{array}
\right ) \left(
\begin{array}{cc}
\frac{\partial \mu }{\partial P} & \frac{\partial \mu }{\partial
X}\\ \frac{\partial {\mathcal M}_{dnkp}}{\partial P} &
\frac{\partial {\mathcal M}_{dmkp} }{\partial X}
\end{array}
\right )
=
\left(
\begin{array}{cc}
\frac{\partial \tilde \mu }{\partial P} & \frac{\partial \tilde
\mu }{\partial X}\\ \frac{\partial \tilde {\mathcal
M}_{dnkp}}{\partial P} & \frac{\partial  \tilde {\mathcal
M}_{dmkp} }{\partial X}
\end{array}
\right ).  \label{mat1}
 \ee

Since the determinant of the first matrix on the left hand side is
$1$ because of (\ref{uni}), the determinants of both hand sides
give
\[
\{\mu, {\mathcal M}_{dmkp} \}=\{ \tilde \mu, \tilde {\mathcal
M}_{dmkp} \}.
\]

One can calculate the left hand side as
\begin{eqnarray*}
\{\mu, {\mathcal M}_{dmkp} \} &=& \frac{\partial \mu}{\partial P}
\frac{\partial {\mathcal M}_{dmkp}}{\partial X}-\frac{\partial
{\mathcal M}_{dmkp}}{\partial P} \frac{\partial \mu}{\partial X},
\\
 &=& \frac{\partial \mu}{\partial P} \left [\left (\frac{\partial M_{dmkp}}{\partial \mu} \right )_{S_{i}(T)\hspace{2mm} fixed} \frac{\partial \mu}{\partial X}+1+\sum_{i=1}^{\infty} \frac{\partial S_{i}(T)}{\partial X}
\mu^{-i} \right] \\
 &-& \frac{\partial \mu}{\partial X}  \left (\frac{\partial M_{dmkp}}
 {\partial \mu} \right )_{S_{i}(T)\hspace{2mm} fixed}  \frac{\partial \mu}{\partial
 P},\\
&=&1+(negative \,\ powers \,\ of \,\  P)
\end{eqnarray*}
where we have used the fact that the terms containing $\left
(\frac{\partial M_{dmkp}}{\partial \mu} \right )_{S_{i}(T)
\hspace{2mm} fixed} $ in the last line cancel.
 Moreover, the Laurent expansions of $\tilde \mu $ and $ \tilde {\mathcal M}_{dmkp}$
contain only non-negative powers of $P$ because of the functional
equations (\ref{fun}). Therefore strictly negative powers of $P$
in the last line should be absent, thus
\begin{equation}
\{\mu, {\mathcal M}_{dmkp} \}=\{\tilde \mu, \tilde {\mathcal
M}_{dmkp} \}=1. \label{can}
\end{equation}

This gives the desired canonical Poisson relation. We now show
that the Lax equation for $\mu$ and ${\mathcal M}_{dmkp}$ are
indeed satisfied. Differentiating equations (\ref{sub}) with
respect to $T_n$ gives
\begin{equation}
\left(
\begin{array}{cc}
\frac{\partial f(\mu, {\mathcal M}_{dmkp})}{\partial \mu} &
\frac{\partial f(\mu, {\mathcal M}_{dmkp})}{\partial {\mathcal
M}_{dmkp}} \\ \frac{\partial g(\mu, {\mathcal M}_{dmkp})}{\partial
\mu} & \frac{\partial g(\mu, {\mathcal M}_{dmkp})}{\partial
{\mathcal M}_{dmkp}}
\end{array}
\right ) \left(
\begin{array}{c}
\frac{\partial \mu }{\partial T_n} \\ \frac{\partial {\mathcal
M}_{dmkp}}{\partial T_n}
\end{array}
\right )
=
\left(
\begin{array}{c}
\frac{\partial \tilde \mu }{\partial T_n}  \\ \frac{\partial
\tilde {\mathcal M}_{dmkp}}{\partial T_n}
\end{array}
\right ).  \label{mat2}
\end{equation}

Combining equations (\ref{mat1}) and (\ref{mat2}), one can
eliminate the derivative matrix of $(f,g)$ by $(\mu, {\mathcal
M}_{dmkp})$ and obtain the matrix relation
 \[
\left(
\begin{array}{cc}
\frac{\partial \mu }{\partial P} & \frac{\partial \mu }{\partial
X}\\ \frac{\partial {\mathcal M}_{dnkp}}{\partial P} &
\frac{\partial {\mathcal M}_{dmkp} }{\partial X}
\end{array}
\right )^{-1} \left(
\begin{array}{c}
\frac{\partial \mu }{\partial T_n} \\ \frac{\partial {\mathcal
M}_{dmkp}}{\partial T_n}
\end{array}
\right ) = \left(
\begin{array}{cc}
\frac{\partial \tilde \mu }{\partial P} & \frac{\partial \tilde
\mu }{\partial X}\\ \frac{\partial \tilde {\mathcal
M}_{dnkp}}{\partial P} & \frac{\partial  \tilde {\mathcal
M}_{dmkp} }{\partial X}
\end{array}
\right )^{-1} \left(
\begin{array}{c}
\frac{\partial \tilde \mu }{\partial T_n}  \\ \frac{\partial
\tilde {\mathcal M}_{dmkp}}{\partial T_n}
\end{array}
\right ).  \label{mat3}
 \]

Since the the determinants of the $2 \times 2$ matrices on both
sides are $1$ because of (\ref{can}), the inverse can also be
written explicitly. In components, thus, the above matrix relation
gives
 \begin{eqnarray}
 \frac{\partial {\mathcal M}_{dmkp}}{\partial X} \frac{\partial
 \mu}{\partial T_n} -\frac{\partial \mu}{\partial X} \frac{\partial
 {\mathcal M}_{dmkp}}{\partial T_n} &=& \frac{\partial \tilde
 {\mathcal M}_{dmkp}}{\partial X} \frac{\partial \tilde \mu}{\partial
 T_n} -\frac{\partial \tilde \mu}{\partial X} \frac{\partial \tilde
 {\mathcal M}_{dmkp}}{\partial T_n}, \no \\
 \frac{\partial
 {\mathcal M}_{dmkp}}{\partial P} \frac{\partial \mu}{\partial T_n}
 -\frac{\partial \mu}{\partial P} \frac{\partial {\mathcal
 M}_{dmkp}}{\partial T_n} &=& \frac{\partial \tilde {\mathcal
 M}_{dmkp}}{\partial P} \frac{\partial \tilde \mu}{\partial T_n}
 -\frac{\partial \tilde \mu}{\partial P} \frac{\partial \tilde
 {\mathcal M}_{dmkp}}{\partial T_n}.
 \label{equal}
 \end{eqnarray}

The left hand side of equation (\ref{equal}) can be calculated
just as we have done above for derivatives in $(P, X)$. For the
first equation of (\ref{equal}),
\begin{eqnarray}
& &\frac{\partial {\mathcal M}_{dmkp}}{\partial X} \frac{\partial
\mu}{\partial T_n} -\frac{\partial \mu}{\partial X} \frac{\partial
{\mathcal M}_{dmkp}}{\partial T_n}  \no \\ &=&\left [\left
(\frac{\partial {\mathcal M}_{dmkp}}{\partial \mu} \right
)_{S_{i}(T)\hspace{2mm} fixed} \frac{\partial \mu}{\partial
X}+1+\sum_{i=1}^{\infty} \frac{\partial S_{i}(T)}{\partial X}
\mu^{-i} \right] \frac{\partial \mu}{\partial T_n}  \no \\ &-&
\frac{\partial \mu}{\partial X} \left [\left (\frac{\partial
{\mathcal M}_{dmkp}}{\partial \mu} \right )_{S_{i}(T)\hspace{2mm}
fixed} \frac{\partial \mu}{\partial T_n}+ n \mu^{n-1}
+\sum_{i=1}^{\infty} \frac{\partial S_{i}(T)}{\partial X} \mu^{-i}
\right], \no
\end{eqnarray}
and terms containing $\left (\frac{\partial M_{dmkp}}{\partial
\mu} \right )_{S_{i}(T)\hspace{2mm} fixed} $ cancel. Thus,
\[
\frac{\partial {\mathcal M}_{dmkp}}{\partial X} \frac{\partial
\mu}{\partial T_n} -\frac{\partial \mu}{\partial X} \frac{\partial
{\mathcal M}_{dmkp}}{\partial T_n} = -\frac{\partial (\mu^n)_{\geq
1}}{\partial X} +(powers \,\ of \,\  P \leq 0). \label{rest1}
\]
By the functional equations (\ref{fun}), we know that the right
hand side of the first equation of (\ref{equal}) has Laurent
expansion with only powers  of $\geq 1$. Therefore only powers of
$P \geq 1$ should survive. Hence
\begin{equation}
\frac{\partial {\mathcal M}_{dmkp}}{\partial X} \frac{\partial
\mu}{\partial T_n} -\frac{\partial \mu}{\partial X} \frac{\partial
{\mathcal M}_{dmkp}}{\partial T_n} = -\frac{\partial (\mu^n)_{\geq
1}}{\partial X}=-\frac{\partial {\mathcal Q}_n}{\partial X}.
\label{fina1}
\end{equation}
For the second equation of (\ref{equal}), we have similarly
\begin{eqnarray*}
\frac{\partial {\mathcal M}_{dmkp}}{\partial P} \frac{\partial
\mu}{\partial T_n} -\frac{\partial \mu}{\partial P} \frac{\partial
{\mathcal M}_{dmkp}}{\partial T_n} &=& -\frac{\partial
(\mu^n)_+}{\partial P} +(negative \,\ powers \,\ of \,\  P ),
\\ &=& -\frac{\partial (\mu^n)_{\geq 1}}{\partial P} +(negative
\,\ powers \,\ of \,\  P ) .
\end{eqnarray*}
By the functional equations (\ref{fun}), noticing the partial
derivative $\partial / \partial P $, we see that  the right hand
side of the second equation of (\ref{equal}) have Laurent
expansion with only nonnegative powers of $P$. Hence only
nonnegative powers of $P$ should survive.

Thus
\begin{equation}
\frac{\partial {\mathcal M}_{dmkp}}{\partial P} \frac{\partial
\mu}{\partial T_n} -\frac{\partial \mu}{\partial P} \frac{\partial
{\mathcal M}_{dmkp}}{\partial T_n} = -\frac{\partial (\mu^n)_{\geq
1}}{\partial P}=-\frac{\partial {\mathcal Q}_n}{\partial P}.
\label{fina2}
\end{equation}

Using (\ref{can}), equations (\ref{fina1}) and (\ref{fina2}) can
be readily solved:
\begin{eqnarray*}
\frac{\partial \mu}{\partial T_n} &= & -\frac{\partial
\mu}{\partial P}\frac{\partial {\mathcal Q}_n}{\partial X} +
\frac{\partial \mu}{\partial X}\frac{\partial {\mathcal
Q}_n}{\partial P}=\{{\mathcal Q}_n, \mu \}, \\ \frac{\partial
{\mathcal M}_{dmkp}}{\partial T_n} &= & -\frac{\partial {\mathcal
M}_{dmkp}}{\partial P}\frac{\partial {\mathcal Q}_n}{\partial X} +
\frac{\partial {\mathcal M}_{dmkp}}{\partial X}\frac{\partial
{\mathcal Q}_n}{\partial P}=\{{\mathcal Q}_n, {\mathcal M}_{dmkp}
\}.
\end{eqnarray*}
This completes the theorem. $\Box$
\section{concluding remarks}

We have studied the Miura map between the dKP and dmKP
hierarchies. We show that the Miura map not only preserves the Lax
formulation of these two hierarchies but also is a canonical map
in the sense that the bi-Hamiltonian structure of the dmKP
hierarchy is mapped to the bi-Hamiltonian structure of the dKP
hierarchy. We further use the twistor construction developed by
Takasaki and Takebe to investigate the solution structure of the
dmKP hierarchy.

 In spite of the results obtained in the paper,
 there are some related problems deserve further investigations.
 We list some of them in the following. \\
 (1) In \cite{ch}, it is shown that the second
Hamiltonian structure $\Theta^{(2)}$ of dKP has
  free field realizations. Since the Miura map is canonical, this suggests the
  possibility of free field realizations of second Hamiltonian structure $J^{(2)}$
  of dmKP \cite{ct}.  \\
 (2) In \cite{du2}, we know that bi-Hamiltonian structure of
Dubrovin-Novikov (DN) type \cite{dn} has geometric structure of
Frobenius manifold \cite{du1}. A natural question is : what's the
geometric meaning of the Miura map between bi-Hamiltonian
structures of DN type? \\
 (3) The dmKP theory should be
investigated without using Miura map. The quasi-classical
$\tau$-function for dKP has been established in
\cite{Ta2,kr1,kr2}. The basic question for dmKP theory is : Does
the quasi-classical $\tau$-function theory exist ? We notice that
the Hirota bilinear equations for KP and mKP are essentially
different \cite{jm}. Also, in \cite{Car2}, the dispersionless
Hirota equation for dKP is obtained. Is there an analogue for dmKP
?\\

{\bf Acknowledgements\/}\\ We would like to thank Prof. J.C Shaw
for useful discussions. JHC thanks for the support of the Academia
Sinica and MHT  thanks for the support of  the National Science
Council of Taiwan under Grant No. NSC 89-2112-M194-018.

\end{document}